\documentclass[fleqn,twoside]{article}
\usepackage{espcrc2}

\usepackage{graphicx}
\usepackage[figuresright]{rotating}
\newcommand{\AmS}{{\protect\the\textfont2
  A\kern-.1667em\lower.5ex\hbox{M}\kern-.125emS}}

\title{Online monitoring system and data management for KamLAND
       }

\author{M. Motoki\address{Research Center for Neutrino Science, 
        Tohoku University, \\ 
        Sendai 980-8578, Japan}%
\thanks{
{\it Corresponding author.}
}
, 
F. Suekane\addressmark, 
K. Tada\addressmark,
 and Y. Tsuda\addressmark
}
       
\begin{document}

\begin{abstract}
%These pages provide you with an example of the layout and style for
%100\% reproduction which we wish you to adopt during the preparation of
%your paper. This is the output from the \LaTeX{} document class you
%requested.
In January 22, 2002, KamLAND started the data-taking.  The KamLAND
detector is a complicated system which consists of liquid scintillator,
buffer oil, spherical balloon and so on. In order to maintain the
 detector safety,
we constructed monitoring system which collect detector status
 information such as
balloon weight, liquid scintillator oil level and so on. 
In addition, we constructed continuous Rn monitoring system for the $^7$Be
 solar neutrino detection.
The KamLAND monitoring system
consists of various network, LON, 1-Wire, and TCP/IP,
and these are indispensable for continuous experimental data acquisition.
%and we can see the condition
% of KamLAND detector from PC via TCP/IP.  
%The physics-data size
%is about 120GB/day and it is stored on LTO tape.The event reconstructed
%data set which is used by physics study is about 100MB/day.  
\vspace{1pc}
\end{abstract}

% typeset front matter (including abstract)
\maketitle

\section{Introduction}

%Text should be produced within the dimensions shown on these pages:
%each column 7.5 cm wide with 1 cm middle margin, total width of 16 cm
%and a maximum length of 19.5 cm on first pages and 21 cm on second and
%following pages. The \LaTeX{} document class uses the maximum stipulated
%%length apart from the following two exceptions (i) \LaTeX{} does not
%begin a new section directly at the bottom of a page, but transfers the
%heading to the top of the next page; (ii) \LaTeX{} never (well, hardly
%ever) exceeds the length of the text area in order to complete a
%section of text or a paragraph. Here are some references:
%\cite{Scho70,Mazu84}.

%KamLAND実験は2002年の1月22日より実験を開始し、これまでに原子炉反ニュート
%リノ欠損及び太陽などからの反ニュートリノの高精度な探索を行った。
%kamLAND測定器は非常に複雑な測定器であり、その状態のモニターは非常に
%じゅうようである。
%次の実験として7BE太陽ニュートリノの検出がある。
%実験で安全にデータをとるためには、監視モニターシステムが必要となる。
%この論文では、モニターシステム、ラドン濃度観測システムについて述べる。

The solar neutrino deficit was a longstanding unsolved problem for
almost 30 years.
%However the recent solar neutrino experiment, Sudbury Neutrino
%Observatory (SNO) detected the evidence
%that neutrino flavor transition caused this deficit~\cite{SNO}. 
%The next step in experiments is to determine the oscillation parameter.
%Based on the solar neutrino measurements to date, one of the most
%plausible parameters with the hypothesis of two-flavor oscillations
%stands at $sin^2 2\theta \sim 0.83$ and $\Delta m^2 \sim 5.5 \times
%10^{-5} eV^2$, which commonly referred to as the large mixing angle
%solution, LMA.
The
remaining solutions were limited to only two, LMA 
 and LOW.
%the former being a particularly promising answer. 
%Hence, the next step was
%to distinguish between LMA and LOW and, finally, to determine the
%oscillation parameters. 
%
The LMA parameter was accessible by a laboratory test based on reactor
neutrinos with a long-baseline of more than 100 km.
KamLAND (Kamioka Liquid scintillator Anti-Neutrino Detector) is such the
experiment aiming to examine the oscillation parameters around LMA~\cite{kamland}.
%The detector and underground facility construction took about 5 years.
%The KamLAND collaboration consists of global institute, Japan, US and China.
The data-taking started in January 22, 2002.
The first results of KamLAND came from a study of reactor anti-neutrino
oscillations based on an analysis of 162 ton-yr exposure data.
KamLAND found fewer electron anti-neutrino events than expected from standard
 assumptions about the electron anti-neutrino propagation at 99.95\%
 C.L~\cite{klpaper01}.
%In a 162 ton-yr exposure the ratio of the number of observed inverse
% $\beta$-decay events to the expected number without electron
% anti-neutrino disappearance was $0.611 \pm 0.085(stat.) \pm 0.041(sys.)$
%for electron anti-neutrino energies $>$ 3.4 MeV.
In the context of two-flavor neutrino oscillations with CPT invariance,
 all solutions to the solar neutrino problem except for the
 ``LMA''region were excluded.
In addition, we searched $\bar{\nu}_e$'s in 
the energy range 8.3~MeV~$<$ E$_{\bar{\nu}_e}$ $<$~14.8 MeV.
No candidates were found for an expected background of $1.1{\pm}0.4$ events.
This result can be used to obtain a limit on $\bar{\nu}_{e}$ fluxes from any
origin~\cite{klpaper02}.

KamLAND detector is complicated system that consists of liquid
scintillator, buffer oil and balloon and so on.
%In order to ta
It is important to take data keeping detector stable.
Therefore we constructed KamLAND detector monitoring system.
We are planning to take $^{7}$Be solar neutrino as a next step.
So, we have to remove background and measure it precisely.
Therefore we constructed continuous radon(Rn) monitoring network system.
% and developing high
%sensitive radon monitor.
%
In this paper, we explain monitoring system that are maintaining the
safety of the KamLAND detector.

%\subsection{KamLAND experiment and detector}
\section{KamLAND detector}

%We normally recommend the use of 1.0 (single) line spacing. However,
%when typing complicated mathematical text \LaTeX{} automatically
%increases the space between text lines in order to prevent sub- and
%superscript fonts overlapping one another and making your printed
%matter illegible.

\begin{figure}[htb]
%\begin{center}
%\includegraphics[angle=270,width=14cm]{xdetector5.epsi}
%\includegraphics[angle=270,width=18pc]{xdetector5.epsi}
\includegraphics[angle=270,width=17pc]{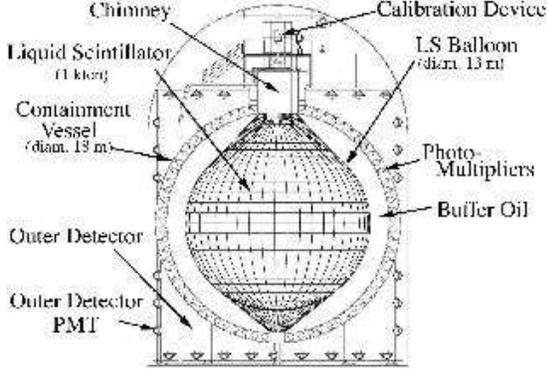}
%\end{center}
\caption{
Schematic diagram of the KamLAND detector.
}
\label{fg:xdetector5}
\end{figure}

The KamLAND detector~(Figure~\ref{fg:xdetector5}) was designed for a low energy neutrino detection
with a large volume, 1000 ton liquid scintilator.
%
%
% The detector is located in the cavern in which Kamiokande used to be. 
There is 2,700mwe of rock overburden. The cosmic-ray background is
reduced to a factor $10^{-5}$.
% compared to that on the earth surface, 
%resulting in 0.34~muons/s for the KamLAND detector volume. 
%The neutron and spallation backgrounds caused by cosmic ray muons was
%reduced to negligible level as described later. 
The core part of the detector is the large volume liquid scintillator(LS). 
The total volume of the LS is $1,200m^3$. 
The liquid scintillator is mixture of 80volume\% of dodecane(C$_{12}$H$_{26}$) plus 
20volume\% of pseudocumen(1,2,4-Trimethylbenzene) plus 1.5g/liter of
PPO(2,5-Diphenyloxazole:C$_{15}$H$_{11}$NO) as a fluor. 
The direct light output of the LS is more than 50\% anthracene
equivalent and the transparency is more than 9m for 400nm wavelength
light.  
The liquid scintillator is viewed by  1,325 17-inch-apparture PMTs
(Hamamatsu R7250) and 554 20-inch-apparture PMTs which are uniformly
distributed in the inner wall of the stainless steel spherical (sss)
tank of diameter 9m.
%This uniform structure helps to reduce systematic error caused by
%non-uniformity of the detector. 
%
%The R7250 PMT was newly developed based of Super Kamiokande 20-inch PMT for 
%KamLAND use, in which clear detection of one photoelectrons and
%good timing resolution are essential. 
%The transit time spread is 2ns and
%the peak to valley ratio is almost 5.   
%
%Currently only the 17-inch PMTs are operational and the light yield is
%about 300
%photo-electrons (pe)/MeV, corresponding energy resolution being   
%$7.5\%/\sqrt{(E(MeV))}$.
The liquid scintillator is contained in a spherical balloon of diameter 13m. 
The balloon is formed by transparent thin plastic films made of  nylon
and EVOH (ethylene vynil alchole).  
The total thickness of the film is 135 $\mu m$ and the light transparency
is 96\% for 400ns wavelength light. 
The balloon is held in a Kevlar net with the mesh size of about 1m
$\times$ 1m at the equator.  
The weight of the balloon is measured by load cell at the top of each Kevlar rope.
The space between the stainless tank and the balloon is filled by
$1,800m^3$ buffer oil (BO), which is a mixture of dodecane and
isoparaffin(C$_{n}$H$_{2n+2}$, $n \sim 14$) oil. 
The isoparaffin oil has larger specific gravity than dodecane. 
The specific gravity of the buffer oil is controlled such a way that BO
is very slightly lighter than LS by adjusting the dodecane/isoparaffin ratio.  
The LS and BO were purified to remove $^{238}$U, $^{232}$Th, $^{40}$K and $^{222}$Rn
before filling in the detector. 
%The purification system makes use of the techniques of water extraction
%and nitrogen purge.
%The water was also purified by using water purification system
%which makes use of RO membrens. 
%The nitrogen was passed through  cooled charcoals to remove Rn before
%the use.    
As a result, $(3.5 \pm 0.5 \times 10^{-18}g/g$ for $^{238}$U, $(5.2 \pm 0.8) \times 10^{-17}g/g $
for $^{232}$Th and  $<2.7 \times 10^{-16}g/g$ for $^{40}$K were achieved. 
%These contamination level are more than enough to perform reactor
%neutrino physics.
%, but not enough for $^7$Be solar neutrino physics.  
%
%The distance between LS and PMT glass is 2m, that between LS and
%stainless steel vessel shell is 2.5m and that between LS and outside
%rock is more than 3 meters. 
%The $\gamma$ ray backgrounds from those materials were effectively
%shielded with the BO and water layer between sss tank and rock and
%are reduced to a negligible level.  
%The Rn produced by PMT glass are blocked by balloon film and 3mm thick 
%acrylic plates at the radius 8.3m from the center. 
%
The region between the sss tank and rock is filled with 3,000$m^3$ of
pure water. 
%The surface of the rock wall is covered with poly-urethan lining to
%prevent  watar leak and Rn coming-in.  
225 20-inch PMTs are submerged in the water and act as cosmic-ray
anti-counter. 
%The surface of the poly-urethan lining wall and sss tank is covered with light
%reflecting sheet (Tyvec) to efficiently catch the Cerenkov light caused
%by cosmic-rays.
% 
The front-end electronics uses Analog Transient Waveform Digitizer
 (ATWD) which captures waveforms from each PMT.
% with effective sampling rate of 1.49ns. 
%Thus pulse shape is recorded in the magnetic tapes and it
%is possible to do full reconstruction of the signals, allowing the
%timing measurement for each single photoelectron signal. 
%
The trigger makes use of Xilinx FPGA and % and it is very flexible. 
%The trigger threshold is 0.7MeV which allows to take $\bar{\nu_e}$ events
%without any loss.  
the global triggers are issued based on number of
hit channels (each channels has about 0.3 p.e. threshold).
% and currently set at 200
%hits ($\sim$ 700 keV) as a prompt trigger and at 120 hits ($\sim$400 keV) as delayed 
%trigger for 1 msec after each prompt triggers. 
%The event rate is $\sim$ 25Hz 
%and the data size amounts to 160 GB/day. 
%This huge data size determines the lowest 
%trigger threshold but it is sufficiently low for the reactor neutrino analysis.
%
Experimental raw data contains trigger information and digital waveforms.
The trigger information contains run number, event number, event time,
and number of hits(PMT).
%PMT signal is sampled by the %Analog Transfer Waveform Digitizer 
%ATWD.
The ATWD simultaneously can capture 4 channels of independent signals at
670MHz sample speeds.
Then it has sliding capture window with 1.49ns step and 1.49x4ns window,
and 128-samples are taken.
The ATWD is equipped with a common-ramp parallel Wilkinson 10-bit ADC.
In order to reduce dead time, the dual-ATWD ping-pong scheme which
requires channel A and B is deployed.
Multiple ATWD channels is used to extend the dynamic range; High,
Midum, and Low gain.
Therefore data size is; 
$2(A and B) \times 3(gains) \times 10(dynamic range) \times
128(samples) = 7.5 kbit/1PMT/1event$.
The status of KamLAND detector (balloon weight, temperature, etc.) are
monitored by using Slow Control Network.
The Slow Control Network consists of LON, CAN and 1-Wire bus.
The CAN bus is used to control the VME crates which include ATWDs.
%data acquisition electronics (ATWD).
The ATWD electronics temperature and electric power consumption and so on are monitored
by this network.
The LON is monitoring network system of all sensors which covers whole
KamLAND area.
%This network is very reliable, the LON network did not have
%any trouble past 3 years.
%This LON network is very reliable.
%the LON network did not have any trouble past 3 years.
The 1-Wire bus is additional network system for Rn monitoring.
The feature of this network is low cost and easy expansion.

\section{LON monitoring system}
% sub section重要なセンサーの位置、種類、目的
% LOnの説明、
\subsection{Sensors}
\begin{figure}[!htb]
%\begin{center}
%\includegraphics[angle=270,width=14cm]{xdetector5.epsi}
\includegraphics[angle=270,width=18pc]{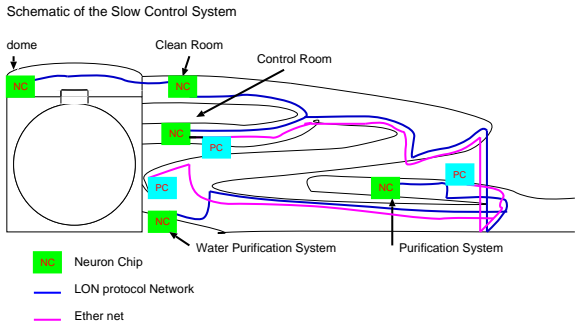}
%\end{center}
\caption{
The schematic view of LON Network in KamLAND area.
}
\label{fg:kamlandarea}
\end{figure}

%
%
%\begin{table*}[htb]
%\begin{table*}[!htb]
%\begin{table*}[!htb]
\begin{table}[!htb]
%\begin{longtable}[htb]
\caption{Sensors and Location}
\label{table:01}
\newcommand{\m}{\hphantom{$-$}}
\newcommand{\cc}[1]{\multicolumn{1}{c}{#1}}
%\renewcommand{\tabcolsep}{2pc} % enlarge column spacing
%\renewcommand{\arraystretch}{1.2} % enlarge line spacing
%\begin{tabular}{@{}llc}
\begin{tabular}{@{}l}
%\begin{tabular*}{\textheight}{@{\extracolsep{\fill}}lllllllllllll}
%\begin{tabular*}{\textheight}{@{}l}
%\begin{longtable}{@{}llc}
\hline
\hline
Sensor (Number of channel) \\
\hline
\hline
Location: Dome area              \\
\hline

                           Load cell (Balloon)  (44) \\
                           Temperature (LS, BOI, BOO) (10) \\
                           Temperature (Anti Water Counter) (4) \\
                           Temperature (Dome)  (1) \\
                           Atmospheric Pressure (Dome)  (1) \\
                           Oxygen monitor (LS, BOI-Dome)  (2) \\
                           N$_2$ Differential pressure (LS-BOI)  (1) \\
                           N$_2$ Differential pressure (LS-Dome)  (1) \\
                           N$_2$ Differential pressure (BOI-Dome)  (1) \\
                           Anti Counter Water Pressure  (2) \\
%Clean room (b)            & \m0.259 & \m0.268 \\
%Control room (c)          & \m2.32  & \m2.83  \\
\hline
Location: Purification area  \\
\hline
                           LS buffer tank (10m$^3$) oil level   (1) \\
                           LS buffer tank (1m$^3$) oil level   (1) \\
                           LS filter out differential pressure   (1) \\
                           LS filter in differential pressure   (1) \\
                           LS Nitrogen supply tank pressure   (1) \\
                           MO buffer tank (1m$^3$) oil level  (1) \\
                           MO buffer tank (15m$^3$) oil level   (1) \\
                           MO filter out differential pressure   (1) \\
                           MO filter in differential pressure   (1) \\
                           MO Nitrogen supply tank pressure   (1) \\
                           Oxigen monitor   (2) \\
                           Frammable gas monitor   (1) \\
                           LS buffer tank (20m$^3$) oil level   (1) \\
                           BO buffer tank (20m$^3$) oil level   (1) \\
                           Detector Oil Level (LS, BOI, BOO)  (3) \\
                           Detector Bottom LS Pressure  (1) \\
                           Detector Bottom Diff. Press. LS-BO   (1) \\
                           Detector Bottom LS flow rate   (1) \\
                           Detector Bottom BOI flow rate   (1) \\
                           Detector Bottom BOO flow rate   (1) \\
\hline
Location: Water Purification area    \\
\hline
                           Final filter out electric conductivity  (1)  \\
                           RO out electric conductivity   (1) \\
                           Temperature (Cooling machine out)   (1) \\
%\hline
%Location: 4th access tunnel      \\
%\hline
                           LS filter differential pressure  (1) \\
                           MO filter differential pressure  (1) \\
\hline
\hline
\end{tabular}\\[2pt]
%\end{tabular*}\\[2pt]
%\end{longtable}\\[2pt]
LS: Liquid Schintilator, \\
BOI: Buffer Oil (Inner), \\
BOO: Buffer Oil (Outer), \\
MO: Mineral Oil (=Buffer Oil).
%\end{table*}
\end{table}
%\end{longtable}
%
%
%
%
Figure~\ref{fg:kamlandarea} shows whole KamLAND area and location of sensors.
Table~\ref{table:01} shows the sensors and locations that are connected
in monitoring system.
%
%
%\clearpage
\subsection{Lon Network and Monitoring system}
The LON was developed by ECHELON Corporation\cite{echelon}.
LonWorks technology is mainly used by building automation networks.
LON network 
consists of the network modules and Neuron chip ICs required to
build intelligent nodes.
The control network configuration is installed in them.
Each LonWorks node includes local processing and I/O to process input
data from sensors and so on.
Each
node includes the capability to communicate with other nodes
because it contains the LONTalk protocol in firmware that are installed
Neuron chip or PROM by using LON builder. 
%\clearpage
The LONTalk
protocol is a complete 7-layer communications protocol that ensures that
nodes can interoperate using an efficient and reliable communications
standard.
%Each node contains Neuron chip, 
%a transceiver to provide the mechanical and electrical interface between
%the node's Neuron chip and the communication media, and
%circuitry to connect the Neuron chip to I/O devices such as sensors.

Each node contains a Neuron chip.
The Neuron chip is a transceiver to provide the electrical I/O interface.
The sensor-data are gathered via these interface.
These data are sent to
another node via LON network.
%\clearpage

%\clearpage

The whole communication length in KamLAND experiment area is about
500~m.
Then we used TP/FT-10 Module Transceiver(Echelon) 
and LW222S LON network
cable (Showa densen).
The maximum network distances is up to 2700~m (Bus topology twisted pair
transceiver), or 500~m (Free topology twisted pair transceiver), we used
bus topology.
The commuication bit rate is 78 kbps.
The TP/FT-10 include Neuron chip TMPN3150(Toshiba), and it need PROM in
order to store the commands.
We attached AT27C256R (Atmel) on TP/FT-10 board.
%In addition, TP/FT-10 has I/O port 
%FTM-10
We made network configuration and each data acquisition program.
Then, these code was written on the PROM that were attached at
TP/FT-10.
\clearpage

\begin{figure}[htb]
%\begin{center}
%\includegraphics[angle=270,width=14cm]{xdetector5.epsi}
\includegraphics[width=18pc]{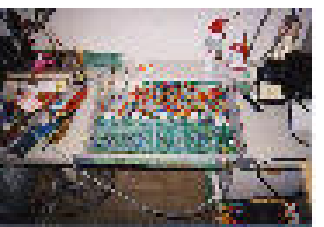}
\includegraphics[width=18pc]{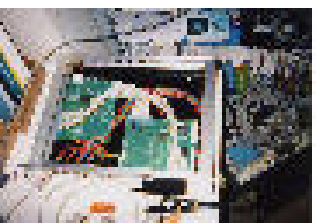}
%\end{center}
\caption{
Uppre figure: ADC Node, Bottom figure: PC interface Node.
}
\label{fg:adcnode}
\end{figure}

\begin{figure}[htb]
%\begin{center}
%\includegraphics[angle=270,width=14cm]{xdetector5.epsi}
\includegraphics[angle=270,width=17pc]{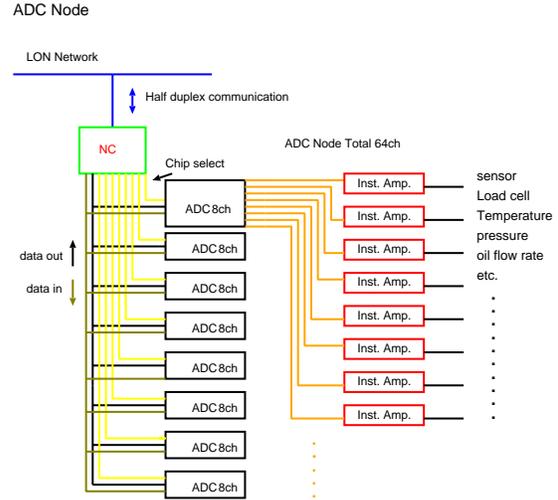}
%\includegraphics[angle=270,width=18pc]{klif2fig.eps}
%\includegraphics[width=18pc]{kldaq1fig-tukau.eps}
%\end{center}
\caption{
The scheme of ADC node with LON Network.
}
\label{fg:scheme01}
\end{figure}

There are two kinds of neuron chip node, ADC data acquisition node and
PC interface node.
Figure~\ref{fg:adcnode} shows ADC data acquisition node and PC interface node.
The schematic view of ADC Node is shown in Figure~\ref{fg:scheme01}.
The ADC data acquisition node consists of ADC MAX186(Maxim) and instrumental
amplifier AD620(Analog Devices).
Almost all sensors have 0-5V or 4-20mA output.
These data are converted by these ADCs.
The ADC boards have 8ch input interfaces and each channel has these instrumental amplifiers (Max
x1000).
Each ADC node can connect 8 ADC boards, hence 1 ADC node can read maximum 64 sensors.

PC interface node gather sensor ADC data via each ADC data acquisition node and
send data to PC via RS232c.
The bit rate of RS232c is set to 115kbps.
%
%\begin{figure}[htb]
%\includegraphics[angle=270,width=15pc]{klif2fig.eps}
%\caption{
%The scheme of LON Network.
%}
%\label{fg:scheme02}
%\end{figure}

\begin{figure}[htb]
%\begin{center}
%\includegraphics[angle=270,width=14cm]{xdetector5.epsi}
%\includegraphics[angle=270,width=18pc]{kladc1fig.eps}
%\includegraphics[angle=270,width=18pc]{klif2fig.eps}
\includegraphics[width=14pc]{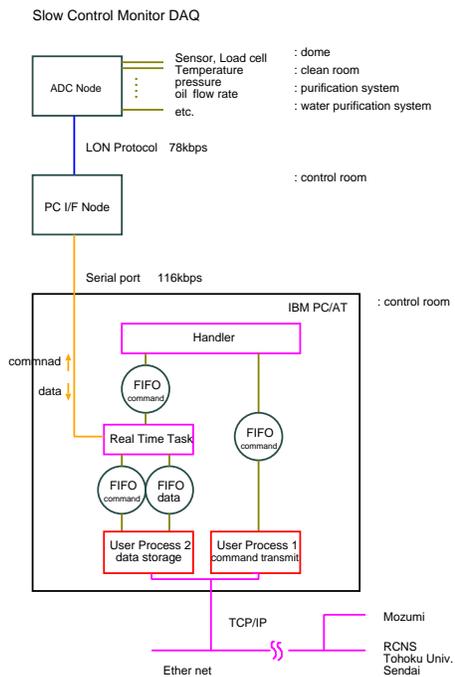}
%\end{center}
\caption{
The scheme of DAQ system with LON Network.
}
\label{fg:scheme03}
\end{figure}

The schematic view of data acquisition is shown in Figure~\ref{fg:scheme03}.
The RT-Linux was used by data acquisition PC. 
These tasks are connected via FIFOs.
User process 1 is human interface of data acquisition system.
Real time task is waked up by handler, and this command is sent from
user process 1.
User process 1 also sends command to each Neuron chip on LON network.
When user process 1 sends command to user process 2 in order to store
monitor data on file, user process 2 starts storing monitor data.
Java GUI interface is deployed for online monitoring of KamLAND area.

%MAX186(maxim), AD620(analog devices)

%\clearpage

\section{1-Wire bus Rn monitoring system}
We constructed continuous $^{222}$Rn monitoring system in the whole KamLAND area.
We used Rn monitor (Model 1027) that was manufactured by Sun Nuclear
Corporation~\cite{Sun}.
In order to gather the $^{222}$Rn density data by using the 1-Wire
network system,
the TTL pulse output read-out interface was attached on Model 1027.

\subsection{$^7$Be Solar Neutrino}
We are planning to detect the $^7$Be Solar Neutrino.
So, we must reduce radio active background, such as $^{210}$Pb, $^{85}$Kr and
so on.
The $^{210}$Pb is serious background to detect the $^7$Be Solar Neutrino and it
is daughter nuclei of the $^{222}$Rn.
The $^{222}$Rn density in the air is very high in the kamioka mine, it is
$\sim$KBq/m$^3$.
%This $^{222}$Rn density is much higher than the Rn density
%obtained at the outside the mine.
This $^{222}$Rn density is much higher than the Rn density that are obtained at the outside of the mine.
In addition, the $^{222}$Rn density in the mine has seasonal variation.
The $^{222}$Rn density is high in summer.
Therefore, at first, we must monitor $^{222}$Rn density in KamLAND area continuously.
Then we must reduce $^{222}$Rn from KamLAND area.

\subsection{1-Wire bus}
The 1-Wire is a bus based on a PC communicating digitally over twisted
pair cable with the 1-Wire components~\cite{1-wire}.
The network is defined with an open drain master/slave multidrop
architecture that uses a resister pull-up to a nominal 5V supply at the
master.
The 1-Wire network consists of a bus master (PC) with controlling software, the
wiring and 1-Wire devces.

The 1-Wire bus was extended in all KamLAND area and Rn monitors were put at
every strategic point.

\subsection{Rn monitoring system}

\begin{figure}[htb]
\includegraphics[width=18pc]{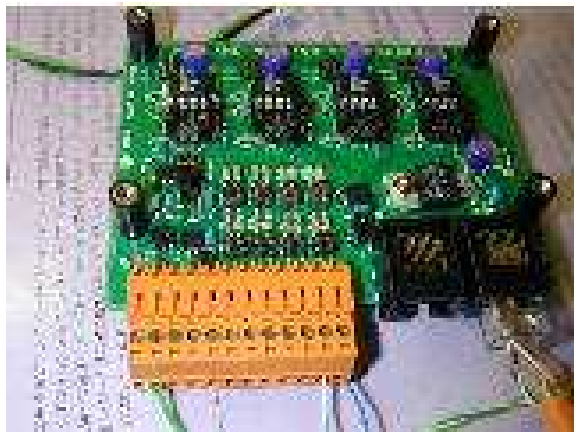}
\caption{
1-Wire Rn monitoring node.
}
\label{fg:1wirekiban}
\end{figure}

Figure~\ref{fg:1wirekiban} shows the Rn monitoring TTL counter node that
include counter, temperature and humidity sensors, and ADCs (4ch).
The TTL pulse signal from Rn counter is counted by DS2423(Maxim).
Then DS2423 send counting data to data acquisition PC.
The efficiency of the Rn detection is influenced by humidity.
Hence, we measure the humidity by using the HIH-3610(Honeywell).
%data acquisition PC via 
This analog data are converted by 
DS2438(Maxim) that include ADC and temperature sensors.
Then these humidity and temperature data are sent to data acquisition PC
via DS2438.
Another sensor-analog data  are converted by 4ch ADC DS2450(Maxim).
Each ADC has instrumentation amprifier AD620(Analog devices), so these
sensor data can be read precisely.

\begin{figure}[htb]
\includegraphics[angle=270,width=18pc]{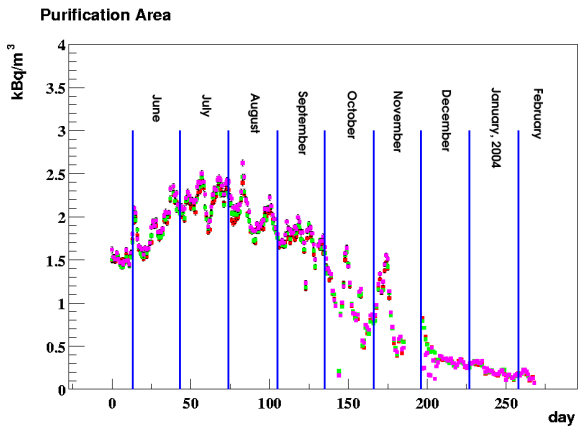}
%\end{center}
\caption{
Seasonal variation of Rn density in KamLAND area.
}
\label{fg:Rn-fig}
\end{figure}

These Rn counting nodes were connected by category 6 LAN cable that were
applied in all KamLAND area.
The electrical power that are consumed by each node is supplied via same network cable.
%The location of Rn monitors are; Dome area (glove box), Dome area
%(purification room), Dome area, Control room, N$_2$ purge tower, KamLAND
%experimental hall entrance, purification room (3), 4th access tunnel,
%and water purification area.
The location of Rn monitors are; Dome area, Control room, N$_2$ purge tower, KamLAND
experimental hall entrance, purification room (3), 4th access tunnel,
and water purification area.
Figure~\ref{fg:Rn-fig} shows the seasonal variation of Rn density in
purification area.
This is due to the direction of the air flow in the kamioka mine.

%\clearpage

%\section{R \& D Sensitive radon monitor}

%\section{KamLAND Experimental data flow}
\section{Data management}
The monitor data are stored on local PC located in the control room and are sent to sendai via berkeley
socket continuously.
The data flow rate of monitoring system is about 30MB/day.
The experimental raw data flow rate is about 160GB/day.
The raw digital waveforms are analyzed and converted to time and charge
information.
After waveform analysis, the data size reduces to about 1/15, it is about
12GB/day.
Then we apply event reconstruction process, the data size reduces to about
1/120, the final data rate is about 100MB/day.
%All the data are stored on HSM(about 10TByte), then we can access all
%data easily.

%D論とか

\section{Summary}
KamLAND experiment started the run from January 22 2002.
Experimental data are accumulated safely by using LON monitoring system.
For the $^7$Be solar neutrino detection, we must reduce radio active background.
In order to measure Rn density in KamLAND area continuously, we constructed Rn
monitoring network by using 1-Wire network.
\\

Acknowledgements 

M.~M. sincerely thanks to K.Yamato for suggestion and communication about LON
monitoring system.
We appreciate supports in the manufacturing process of network by
H.Hanada, M.Nakajima, T.Nakajima, T.Takayama and valuable help by
K.Eguchi, K.Furuno, H.Ikeda, K.Inoue, Y.Kishimoto, M.Koga, T.Mitsui, K.Nakamura, J.Shirai, A.Suzuki,
K.Tamae, E.Yakushev, and other members of KamLAND collaboration.

%%%%%%%%%%%%%%謝辞
%やまと、ふるの、はなだ、いけだ、いのうえ、みつい、なかじまみのる、
%なかじまたかし、なかむらけんご、しらい、すずきあつと、たかやま、たまえ
%えぶげに、他KamLANDのみなさん

\end{document}